\documentclass{moriond}

\usepackage{amsmath}
\usepackage{lineno}

\def\Journal#1#2#3#4{{#1} {\bf #2}, #3 (#4)}

\def\PLB{{\em Phys. Lett.}  B}
\def\PRL{\em Phys. Rev. Lett.}
\def\PRD{{\em Phys. Rev.} D}

\def\CS{{\em Comput.Softw.Big Sci.}}
\def\EPJC{{\em Eur. Phys. J.} C}
\def\JHEP{\em J. High Energ. Phys.}
\def\PTP{\em Prog. Theor. Phys.}
\def\PLB{{\em Phys. Lett.} B}

\def\be{\begin{equation}}
\def\ee{\end{equation}}
\def\bea{\begin{eqnarray}}
\def\eea{\end{eqnarray}}

\newcommand{\Photo}{}

\begin{document}
\vspace*{4cm}
\title{Semileptonic $B$ decays from Belle and Belle II}

\author{ Lu Cao (on behalf of the Belle and Belle~II collaborations)}

\address{Deutsches Elektronen-Synchrotron (DESY),\\
 Notkestraße 85, Hamburg, Germanyd}

%\linenumbers

\maketitle

\abstracts{This proceeding summarizes recent measurements of the CKM matrix elements $|V_{cb}|$, $|V_{ub}|$, and lepton-flavor universality tests in semileptonic B decays from the Belle and Belle~II experiments. The decay branching factions $R(D^{*})$ and $R(X)$ are measured with the Belle~II early dataset and the results are found to be consistent with the Standard Model predictions. From a simultaneous measurement of untagged $B^0 \to \pi^{-} \ell^{+} \nu$ and $B^{+} \to \rho^0 \ell^{+} \nu$, the $|V_{ub}|$ value is determined incorporating external theoretical constraints on the decay form factors. The first measurement of the full angular coefficients for $B \to D^* \ell \nu$ decays at Belle provides comprehensive information for determining the decay form factors and $V_{\mathrm{cb}}$. This presentation also includes measurements of the ratios $|V_{ub}|^{\mathrm{excl}}/|V_{ub}|^{\mathrm{incl}}$ and $|V_{ub}|^{\mathrm{incl}}/|V_{cb}|^{\mathrm{incl}}$.
} 

\section{Lepton-flavor universality tests}

\begin{figure}[h!]
    \centering
    \includegraphics[width=0.8\linewidth]{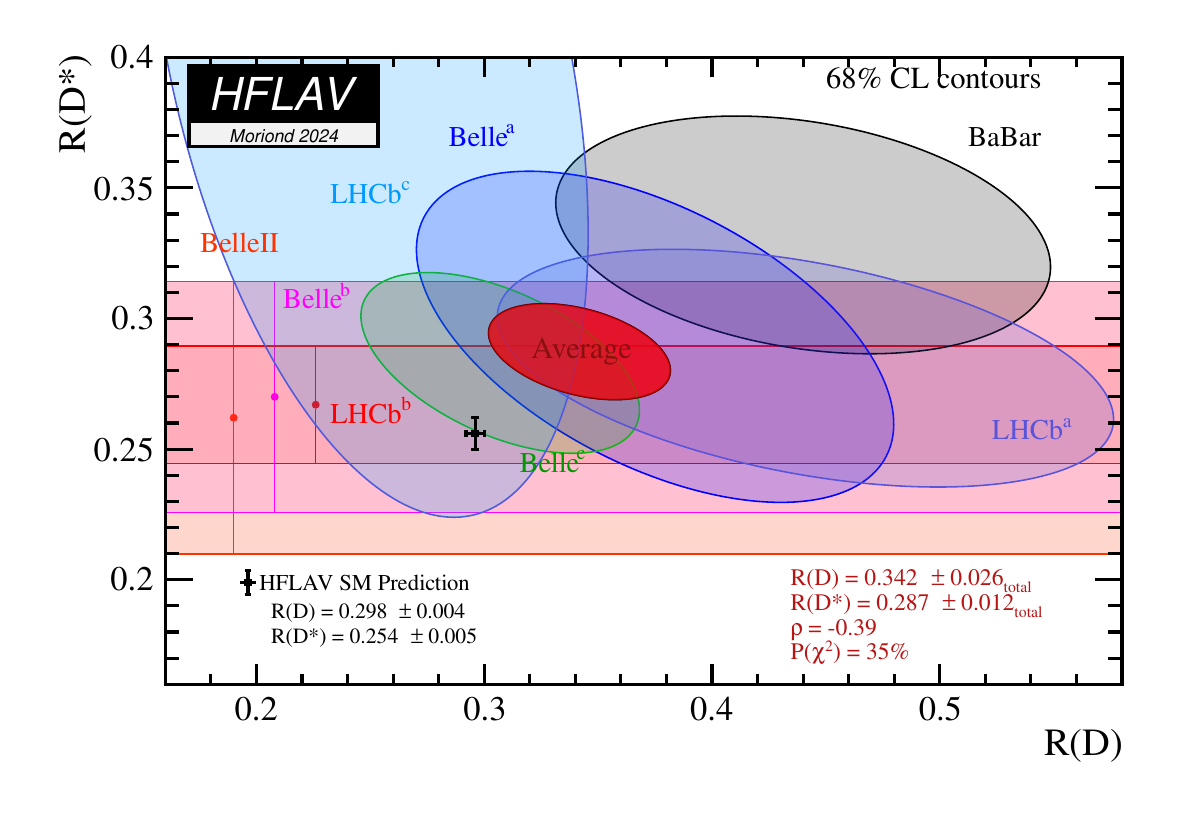}
    \caption{The latest world average results of $R(D)$-$R(D^{*})$ from HFLAV (updated on the 20th of May 2024). %cite{hflav2024}.
    }
     \label{fig:hflav2024}
\end{figure}

In the Standard Model of particle physics (SM), the $W$ boson couples equally to the leptons $\tau$, $\mu$ and $e$. Semileptonic $B$ decays offer stringent tests of this lepton-flavor universality and are thus sensitive to new physics beyond the SM. One of the preferred observables for testing this universality is the ratio of decay branching fractions, $R(H_{{\tau/\ell}})$, which compares decays involving the $\tau$ lepton to those with the lighter leptons $\ell = \mu, e$. This ratio is advantageous because it cancels out normalization effects and correlated uncertainties, providing a more precise test of lepton-flavor universality.

\begin{equation}
R(H_{{\tau/\ell}}) = \frac{\mathcal{B}(B \to H {{ \tau }} \nu)}{\mathcal{B}(B \to H {{ \ell }} \nu)},
\end{equation}
where the final state hadron $H$ can be $D^{(*)}$, $\pi$, or other hadrons from exclusive modes, or it can be an inclusive hadronic system $X$. The latest world average of experimental observations is shown in Fig.~\ref{fig:hflav2024}, revealing a tension with the SM prediction at a level of approximately 3 standard deviations. 

The values of $R(D^{*})$ \cite{kojima} and $R(X)$ \cite{henrik} have been recently measured using a $189\, \mathrm{fb}^{-1}$ sample of electron-positron collision data at the Belle~II experiment. This corresponds to $N_{B \bar{B}}=$ $(198.0 \pm 3.0) \times 10^6 B \bar{B}$ pairs, collected at the $\Upsilon(4S)$ resonance during the 2019-2021 run period.

\subsection{Measurement of $R(D^{*})$ using hadronic B tagging at the Belle II experiment}\label{subsec:RD*}

In this analysis \cite{kojima}, one of the $B$ mesons originating from the $\Upsilon(4S)$ decay is fully reconstructed via hadronic decay modes as $B_{\text{tag}}$ using the official tagging algorithm \cite{fei}. The remaining particles in the event are used to reconstruct the other $B$ meson in the decays $\bar{B} \to D^* \tau^{-} \bar{\nu}_\tau$ and the normalization mode $\bar{B} \to D^* \ell^{-} \bar{\nu}_{\ell}$. Only leptonic $\tau$ decays, $\tau^{-} \to e^{-} \bar{\nu}_e \nu_\tau$ and $\tau^{-} \to \mu^{-} \bar{\nu}_\mu \nu_\tau$ are considered. The modeling of signal efficiency, major backgrounds, and the residual energy deposited in the electromagnetic calorimeter (ECL) $E_{\mathrm{ECL}}$, are calibrated using a data-driven approach and validated in the sideband regions individually. After applying all selections and corrections, the value of $R(D^{*})$ is extracted directly from a two-dimensional fit to $E_{\mathrm{ECL}}$ and the missing mass squared $M_{\text {miss }}^2$,
\begin{equation}
M_{\text {miss }}^2=\left(E_{\text {beam }}^*-E_{D^*}^*-E_{\ell}^*\right)^2-\left(-\vec{p}_{B_{\text {tag }}}^*-\vec{p}_{D^*}^*-\vec{p}_{\ell}^*\right)^2 \text {, } \label{eq:mm2}
\end{equation}
where $E_{\text {beam }}^*=\sqrt{s} / 2$ represents the center-of-mass (c.m.) beam energy whereas $E_{B_{\text {tag }}}^*(\vec{p}_{B_{\text {tag }}^*}^*)$, $E_{D^*}(\vec{p}_{D^*}^*)$, and $E_{\ell}$ $(\vec{p}_{\ell}^*)$ are the energies (momentum three-vectors) of the $B_{\text {tag }}, D^*$, and $\ell$, respectively, in the c.m. frame. The preliminary result is obtained as 

\begin{equation}
R\left(D^{ * }\right) = 0.262 ^{ +0.041 }_{ -0.039 }\text { (stat) }^{ +0.035 }_{ -0.032 }\text{ (syst) ,}
\end{equation}
which is consistent with SM predictions \cite{hflav2023}. This value has been included in the global fit of HFLAV shown in Fig.~\ref{fig:hflav2024}. 

\subsection{First measurement of $R(X)$ as an inclusive test of the $b\to c \tau \nu$ anomaly}\label{subsec:RX}
\begin{figure}[h!]
    \centering
    \includegraphics[width=0.7\linewidth]{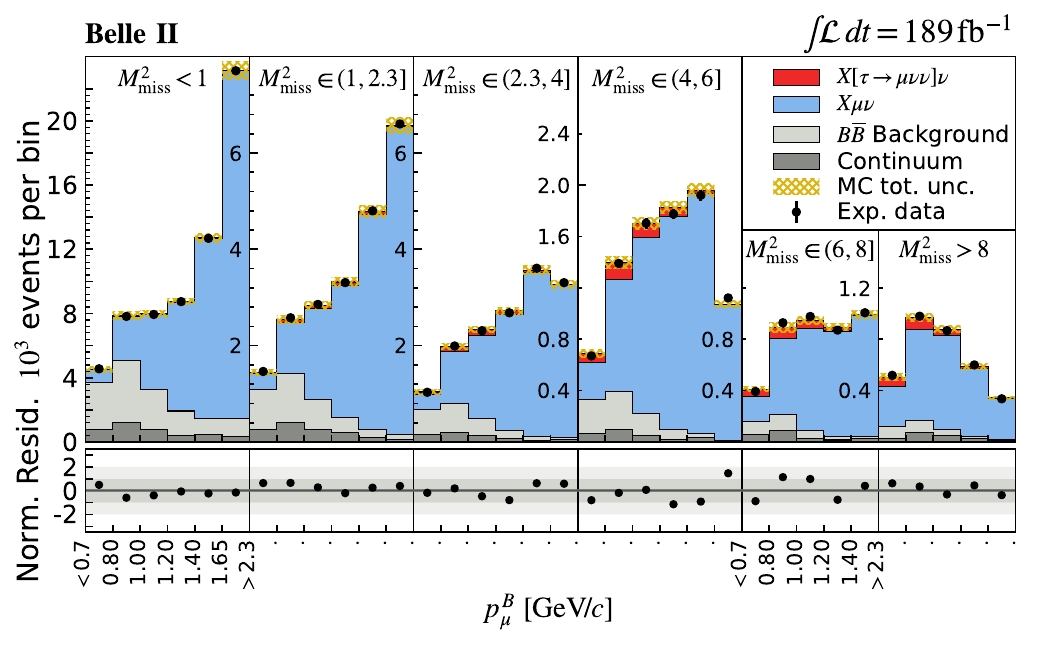}
    \caption{The flattened spectra in intervals of $(M_{\text {miss }}^2 ,p_{\ell}^{B})$ as used in the signal extraction fit after fitting. The intervals of $M_{\text {miss }}^2$ are given in GeV$^{2}/c^{4}$. }
     \label{fig:RX}
\end{figure}

Similar to the $R(D^{*})$ analysis \cite{kojima}, this measurement \cite{henrik} also utilizes the hadronic tagging \cite{fei} method and reconstructs the tauon from its leptonic decays. To reconstruct the inclusive hadronic systems $X$ in $B \to X \tau (\ell) \nu$, all remaining tracks and neutrals not involved in the reconstruction of $B_{\text {tag }}$ or leptons are combined. In order to correct the potential mismodeling of the $B \to X_{c} \ell \nu$ decays, a data-driven reweighting is derived from the control sample. The signal yields are extracted simultaneously for electrons and muons modes using a two-dimensional fit of the missing mass squared $M_{\text {miss }}^2$ and the lepton momentum in B-rest frame $p_{\ell}^{B}$. The post-fit projection for the $\mu$ mode is shown in Fig.~\ref{fig:RX}.

With the fitted signal yields and correlations, the value of $R\left(X_{\tau / \ell}\right)$ and its uncertainty can be calculated via $R\left(X_{\tau / \ell}\right)=$ $\left(N_{\tau \rightarrow \ell}^{\text {meas }} / N_{\ell}^{\text {meas }}\right)\left(N_{\ell}^{\text {sel }} / N_{\tau \rightarrow \ell}^{\text {sel }}\right)\left(N_\tau^{\text {gen }} / N_{\ell}^{\text {gen }}\right)$ using $N_\tau^{\text {gen }}=N_{\tau \rightarrow \ell}^{\text {gen }} / \mathcal{B}(\tau \rightarrow \ell \nu \nu)$ and the appropriate uncertainty propagation. We find $R\left(X_{\tau / \ell}\right)$ for electrons and muons of
\begin{equation}
\begin{aligned}
& R\left(X_{\tau / e}\right)=0.232 \pm 0.020 \text { (stat) } \pm 0.037 \text { (syst), and } \\
& R\left(X_{\tau / \mu}\right)=0.222 \pm 0.027 \text { (stat) } \pm 0.050 \text { (syst), }
\end{aligned}
\end{equation}
respectively. By combining light-lepton flavors in a weighted average of correlated values, we obtain
\begin{equation}
R\left(X_{\tau / \ell}\right)=0.228 \pm 0.016 \text { (stat) } \pm 0.036 \text { (syst). } 
\end{equation}

\begin{figure}[h!]
    \centering
    \includegraphics[width=0.6\linewidth]{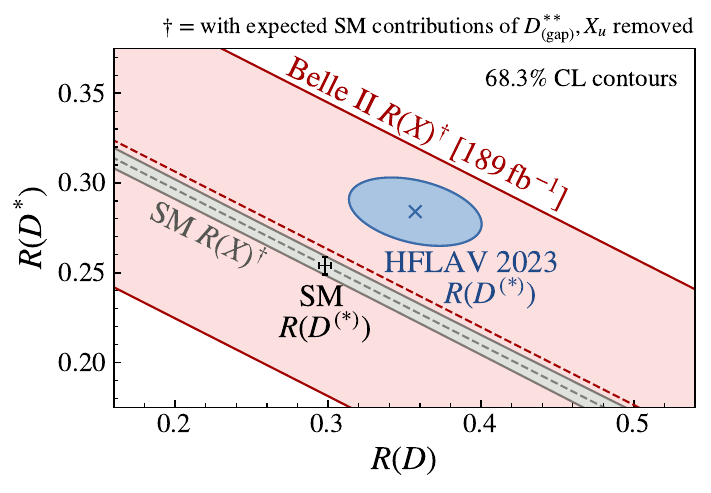}
    \caption{Constraints on $R(D^{(*)})$ from the measured $R(X)$ value (red), compared to the world average (blue) and the SM expectation.
    }\label{fig:RX2}
\end{figure}

The observed result agrees with the SM expectations, e.g. $R(X)_{SM} = 0.221 \pm 0.004$~\cite{RXSM}. After removing the expected contributions of $\mathcal{B}(B \to D^{**}_{(\text{gap})} \tau(\ell) \nu)$ and $\mathcal{B}(B \to X_{u} \tau(\ell) \nu)$ from the measured $R(X)$, the remaining component provides a distinct cross-check of the exclusive $R(D^{(*)})$. As illustrated in Fig.~\ref{fig:RX2}, the result is consistent with the world averaged $R(D^{(*)})$ and also the SM predictions \cite{hflav2023} within the current uncertainty.

\section{CKM matrix elements $|V_{cb}|$ and $|V_{ub}|$}

In the SM, the Cabibbo-Kobayashi-Maskawa (CKM) matrix describes the quark mixing and accounts for $CP-$violation in the quark sector \cite{Cabibbo:1963yz,Kobayashi:1973fv}. One of the crucial tests of the SM is the precise determination of the magnitude of the matrix elements. In $b-$flavor scope, the corresponding world averages of $|V_{xb}|$ from exclusive and inclusive determinations exhibit a disagreement of about 3 standard deviations \cite{hflav2023}. We conducted several measurements with new strategies to further investigate these tensions.

\subsection{Exclusive $|V_{ub}|$ from simultaneous measurements of untagged $B^0 \to \pi^{-} \ell^{+} \nu$ and $B^{+} \to \rho^0 \ell^{+} \nu$ decays}\label{subsec:pirho}

This measurement uses a data sample of 387 million $B \bar{B}$ meson pairs recorded by the Belle II detector at the SuperKEKB electron-positron collider between 2019 and 2022. The signal decays of $B^0 \to \pi^{-} \ell^{+} \nu$ and $B^{+} \to \rho^0 \ell^{+} \nu$ are reconstructed without identifying the partner $B$ meson. The reconstructed events are separated into 13 intervals for the pion mode and 10 intervals for the rho mode of squared momentum transfer $q^{2}$. The signal yields of the two modes are simultaneously extracted from a two-dimensional grid of the energy difference $\Delta E = E^{*}_{B} - E^{*}_{\mathrm{beam}}$ and the beam-constrained mass $M_{bc}=\sqrt{E^{*2}_{\mathrm{beam}} -|\vec{p}^{\,*}_{B}|^{2}}$ in each $q^{2}$ bin. Here, $E^{*}_{\mathrm{beam}}$, $E^{*}_{B}$ and $\vec{p}^{\,*}_{B}$ are the beam energy, reconstructed $B$ energy, and reconstructed $B$ momentum, all determined in the c.m. frame, respectively. With this novel method, cross-feed signals can be properly linked between the two decay modes. 

The partial branching fractions are determined from the fitted signal yields after efficiency corrections as a function of $q^{2}$. Furthermore, the total branching fraction is computed as the sum of these partial branching fractions, accounting for systematic correlations. As preliminary results, we obtain total branching fractions $\mathcal{B}(B^0 \to \pi^{-} \ell^{+} \nu_{\ell})=(1.516 \pm 0.042$ (stat) $\pm 0.059$ (syst) $) \times 10^{-4}$ and $\mathcal{B}(B^{+} \to \rho^0 \ell^{+} \nu_{\ell})=(1.625 \pm 0.079$ (stat) $\pm 0.180$ (syst) $) \times 10^{-4}$. These results are consistent with the world averages, and the precision is comparable to previous measurements from Belle and BaBar.

\begin{figure}[b]
    \centering
    \includegraphics[width=0.45\linewidth]{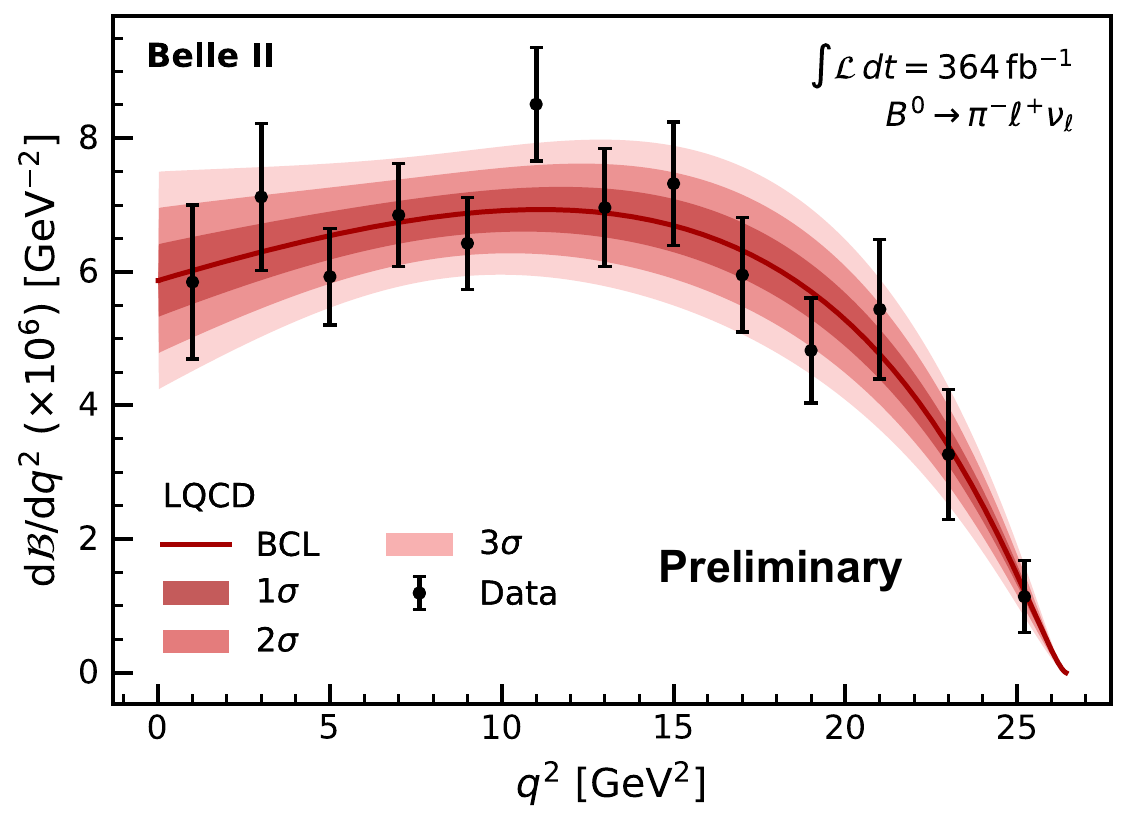}
    \includegraphics[width=0.45\linewidth]{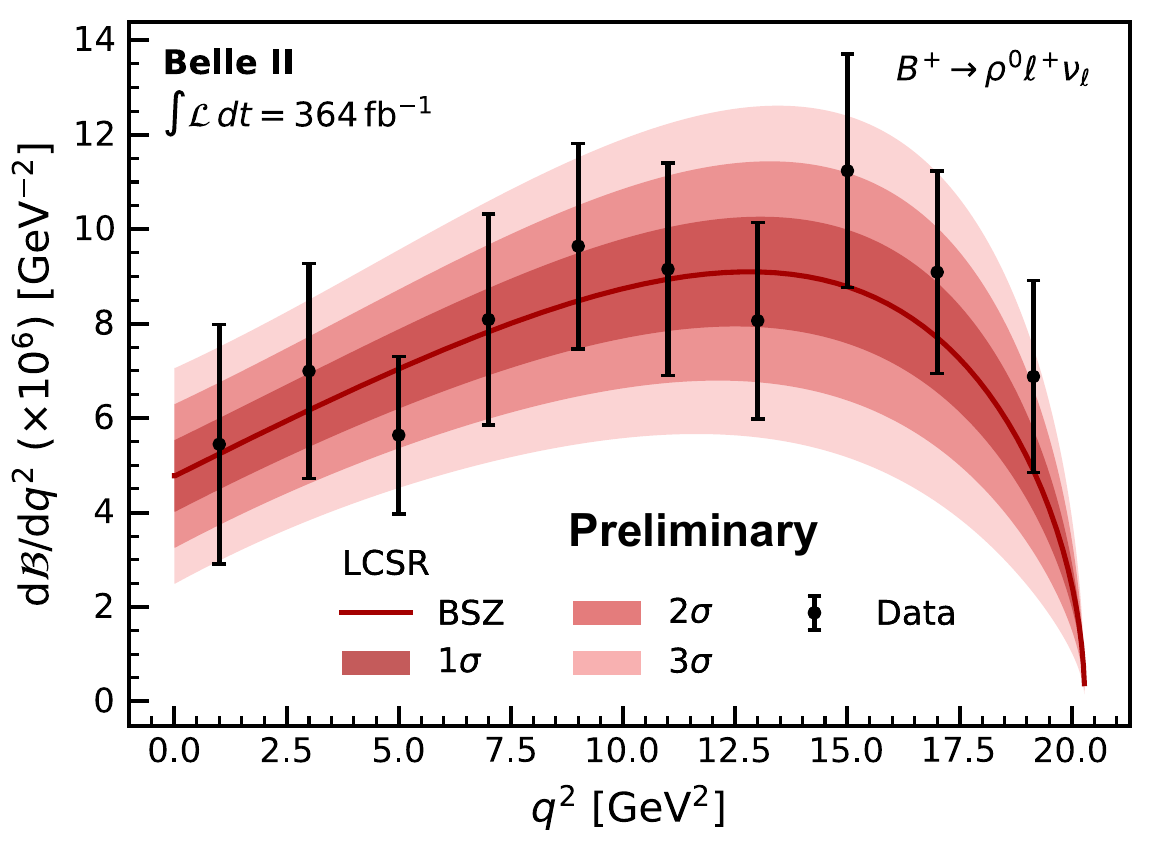}
    \caption{The preliminary results of the differential $q^{2}$ spectra for $B^0 \to \pi^{-} \ell^{+} \nu$ (left) and $B^{+} \to \rho^0 \ell^{+} \nu$ (right) shown with the one, two and three standard-deviation uncertainty bands for fits using constraints on the form factors.
    }\label{fig:pirho}
\end{figure}

For extracting $|V_{ub}|$, the decay form factors of $B^0 \to \pi^{-} \ell^{+} \nu$ are parameterized using the Bourrely-Caprini-Lellouch (BCL) model~\cite{Bourrely:2008za}, and the Bharucha-Straub-Zwicky (BSZ) parametrization \cite{BSZ} is employed for $B^{+} \to \rho^0 \ell^{+} \nu$. By fitting the measured partial branching fractions of $B^0 \to \pi^{-} \ell^{+} \nu$ as functions of $q^2$, and incorporating constraints on non-perturbative hadronic contributions from lattice QCD calculations \cite{flag2023}, we obtain the preliminary result $|V_{u b}|=(3.93 \pm 0.09 \pm 0.13 \pm 0.19) \times 10^{-3}$, where the uncertainties are statistical, systematic, and theoretical, respectively. The preliminary result from the $B^{+} \to \rho^0 \ell^{+} \nu$ decay including the constraints from light-cone sum rule (LCSR) ~\cite{BSZ} is $|V_{u b}|=(3.19 \pm 0.12 \pm 0.17 \pm 0.26) \times 10^{-3}$.  The $|V_{u b}|$ values obtained from the $B^0 \to \pi^{-} \ell^{+} \nu$ mode are consistent with previous exclusive measurements. The result from the $B^{+} \to \rho^0 \ell^{+} \nu$ mode is lower but remains consistent with previous experimental determinations from $B \to \rho \ell \nu$ decays. In both cases, the precision is limited by theoretical uncertainties.

\subsection{Angular coefficients of $B \to D^{*} \ell \nu$ and exclusive $|V_{cb}|$}\label{subsec:vcb}

Using the full Belle sample, comprising $772 \times 10^6$ $B$ meson pairs collected at the $\Upsilon(4S)$ resonance, we performed the first measurement of the complete set of angular coefficients for exclusive $B \to D^* \ell \nu$ decays. The analysis strategy closely follows the methodology outlined in the previous Belle measurement \cite{markusprd}, with modifications to facilitate the measurement of angular coefficients in four bins of the hadronic recoil parameter $w = (m_B^2+m_{D^*}^2-q^2)/2 m_B m_{D^*}$. We separately extract the results for $\ell = e, \mu$ and $B = B^{0}, B^{+}$ modes. In each $w$ bin, the signal yields are determined in bins of the decay angles $\theta_{\ell}, \theta_V$, and $\chi$. $\theta_{\ell}$ is the angle between the lepton and the direction opposite to the $B$ meson in the virtual $W$-boson rest frame, $\theta_V$ is the angle between the $D$ meson and the direction opposite the $B$ meson in the $D^*$ rest frame, and $\chi$ is the angle between the two decay planes spanned by the $W^{+}-\ell$ and $D^*-D$ systems in the $B$ meson rest frame.  

The obtained angular coefficients enable us to determine the form factors describing the $B \to D^*$ transition and the magnitude of $V_{\mathrm{cb}}$. Utilizing various sets of recent lattice QCD calculations for the form factors, we find $\left|V_{\mathrm{cb}}\right|=(41.0 \pm 0.7) \times 10^3$ based on the Boyd-Grinstein-Lebed (BGL) parameterization \cite{BGL}. This result is in agreement with the fit of the one-dimensional differential spectra determined from the same dataset \cite{markusprd} and also with the currently most precise determinations from inclusive $B \to X_c  \ell  \nu$ decays \cite{vcb1,vcb2,vcb3}. Additionally, we investigate potential lepton flavor universality violation as a function of $w$ and analyze the differences in the angular distributions of electrons and muons. No deviation from SM expectations is observed.

\subsection{Simultaneous determination of inclusive and exclusive $|V_{ub}|$}\label{subsec:my}

The first simultaneous determination of $|V_{ub}|$ using inclusive and exclusive decays has been performed at Belle~\cite{cao:2023prl}. The event reconstruction strategies are inherited from the previous Belle $B \to X_u  \ell \nu$ analysis \cite{cao:2021prd}. To distinguish exclusive $B \to \pi  \ell \nu$ decays from other inclusive $B \to X_u  \ell \nu$ events and backgrounds, we employ a two-dimensional fit of $q^2$ and the number of charged pions in the hadronic $X_u$ system. The $B \to \pi \ell \nu$ form factors are parameterized with the BCL expansion \cite{Bourrely:2008za} and constrained to the LQCD calculations \cite{flag2023} or the combined global fit \cite{flag2023} of previous experimental observations and LQCD. With the nominal setup incorporating the constraints based on the full theoretical and experimental knowledge of the $B \to \pi  \ell \nu$ form factor shape, we obtain $\left|V_{ub}^{\mathrm{excl.}} \right| = (3.78 \pm 0.23 \pm 0.16 \pm 0.14)\times 10^{-3}$ and $\left|V_{ub}^{\mathrm{incl.}} \right| = (3.88 \pm 0.20 \pm 0.31 \pm 0.09)\times 10^{-3}$ with the uncertainties being the statistical, systematic, and theoretical errors. The ratio $\left|V_{ub}^{\mathrm{excl.}} \right| / \left|V_{ub}^{\mathrm{incl.}} \right| = 0.97 \pm 0.12$ is found to be compatible with unity. Moreover, the averaged $|V_{ub}|$ derived from the inclusive and exclusive determinations incorporating LQCD and additional experimental information, is $(3.84 \pm 0.26)\times 10^{-3}$. This result is in agreement with the expectation from CKM unitarity \cite{CKMfitter2021} of $|V_{ub}^{\mathrm{CKM}}| = (3.64 \pm 0.07)\times 10^{-3}$ within 0.8 standard deviations.

\subsection{Ratio of inclusive $|V_{ub}|$ and $|V_{cb}|$}\label{subsec:ratio}

The semileptonic inclusive decays $B \to X_u  \ell \nu$ and $B \to X_c  \ell  \nu$ are analyzed using the full Belle sample, employing the Belle II hadronic tagging algorithm \cite{fei}. The $B \to X_u  \ell  \nu$ signal yields are extracted through a two-dimensional fit on $q^{2}$ and the charged lepton energy in the B-meson rest frame $p_{\ell}^{B}$. Meanwhile, $B \to X_c  \ell  \nu$ yields are obtained by subtracting contributions from other decays in the total $B \to X \ell \nu$ sample. This measurement focuses on the partial phase space region with $p_{\ell}^{B} > 1 \, \mathrm{GeV}$, known for cleaner experimental backgrounds in $B \to X_{u}\ell\nu$ decays. The preliminary result for the partial branching faction ratio is $\Delta\mathcal{B}(B \to X_{u}\ell\nu)/\Delta\mathcal{B}(B \to X_{c}\ell\nu) = 1.96(1 \pm 8.4\% {\text{(stat)}} \pm 7.9\% {\text{(syst)}})\times 10^{-2}$. This ratio provides insight into the inclusive $|V_{ub}|/|V_{cb}|$ ratio, incorporating theoretical inputs of partial decay rates for both decays. The obtained $|V_{ub}|/|V_{cb}|$ value is consistent with the world averages \cite{hflav2023}. Furthermore, by taking the external normalization of $\Delta \mathcal{B}(B \to X_c \ell  \nu)$, the resulting $|V_{ub}|$ is found to be in good agreement with previous Belle measurement~\cite{cao:2021prd}.

\section{Summary}

The Belle and Belle~II experiments have recently provided many new results in semileptonic $B$ decays. The measured $R(D^{*})$ and $R(X)$ for testing the lepton-flavor universality are found to be consistent with the SM predictions. The long-standing "$V_{xb}$ puzzle" remains unresolved, although the recent determinations line towards improved agreement between exclusive and inclusive decays. Continued efforts in both experimental and theoretical realms are essential. For instance, some of the experimental uncertainties are expected to be reduced with more collected data at Belle~II, and the ongoing developments in theoretical studies are anticipated to further refine experimental simulations. Moreover, beyond these important results, the accumulated knowledge on MC modeling and validated novel approaches will be beneficial for future measurements.

\end{document}